\documentclass[12pt]{article}
\usepackage{epsfig,amsfonts}

\newfam\mbffam \def\mbf#1{{\mathbf#1}}
\def\avg#1{\langle #1 \rangle} 
 		\def\tr{\mathrm{tr}}

 \def\Vp{\mbf p} \def\Vx{\mbf x}\def\VA{\mbf A}

\newcommand{\sheaf}[1]{\ensuremath{\mathcal{#1}}}
 
\newcommand\CD{\sheaf{D}}

\newcommand\CL{\sheaf{L}}

\newcommand\CV{\sheaf{V}}

\begin{document}
\begin{flushright}
UTTG-12-98
\end{flushright}
\begin{center}
{\Large\bf  Parity breaking and phase transition induced by 
a magnetic field in high $T_c$ superconductors}

\vskip 2cm
{W. Vincent Liu\footnote{liu@physics.utexas.edu} \\
\it Theory Group, Department of Physics, University of Texas, Austin, Texas
78712-1081} 

\vskip 3cm
\begin{abstract}
We suggest that the $2+1$ dimensional Gross--Neveu model can give 
an effective field theory description of low-energy quasiparticles in
high temperature superconductors.  
The magnetic catalysis of dynamical symmetry breaking is examined. The
model shows that a magnetic field can induce a phase transition
associated with parity breaking. In particular, it is intended to
give an explanation of 
a second phase transition observed in a recent experiment
by Krishana {\it et al.}

\end{abstract}
\end{center}
\newpage

\section{Introduction}

The catalysis of chiral symmetry breaking by a magnetic field was
studied in Nambu--Jona-Lasino models in $2+1$
\cite{Gusynin+:94+95,Krive+:92,Klimenko:92} and $3+1$
\cite{Gusynin+:95:plb} dimensions. It was also extended to the case
of QED, external non-Abelian chromomagnetic fields and
finite temperatures, curved space, supersymmetry,
etc. \cite{Gusynin:97:talk},  confirming the universality of the
phenomenon.   By far it is already well known  that a magnetic field
is a strong catalyst of dynamical symmetry breaking, leading to
generating a fermion dynamical mass even at the weakest attractive
interactions between fermions.

In this paper we promote the hypothesis that this magnetic catalysis
of dynamical symmetry breaking can be used to explain the results of  
a recent experiment on Bi$_2$Sr$_2$CaCu$_2$O$_8$ (BSCCO) by Krishana
{\it et al.} \cite{Krishana+:97}. At temperature well below the
superconducting critical temperature ($T_c$), 
they discovered  a kink-like feature of 
thermal conductivity at a transition magnetic field, which depends on
temperature approximately in a quadratic form. They suggested 
 that the field induced a second phase
transition in the superconducting state:
a large gap appeared
abruptly and time reversal symmetry was broken.  
By now, several possible mechanisms of this phenomenon
have been discussed
\cite{Laughlin:98,+Mavromatos:97,Ramakrishnan:98,
Semenoff+:98}.  Laughlin~\cite{Laughlin:98} argued
that a small parity-violating $d_{xy}$ superconducting order
parameter was developed, and his hypothesis led to a model free-energy
functional 
which could display a phase transition induced by a magnetic
field. But he did not tell how
the $d_{xy}$  order parameter was developed.  

In the following, we shall argue that  an extended version of the
Gross--Neveu model 
\cite{Gross-Neveu:74,Rosenstein-Warr-Park:91,Weinberg:97} in $2+1$
dimensions is suitable for studying the low energy excitations of high
temperature superconductors. The theory shows that an
external magnetic field $B$ can induce a
continuous phase transition at  temperatures below the superconducting
critical temperature: an additional
gap is developed and parity is broken. The temperature
dependence of the critical magnetic field shall be calculated
and fitted into the experiment of Krishana {\it et
al.}\cite{Krishana+:97} Unlike those parity-conserving
gauge models studied in
Refs.~\cite{+Mavromatos:97,Semenoff+:98}, 
parity is  broken in our model.  Since parity violation is
becoming popular in studying high $T_c$ superconductors (see
e.g. \cite{Rokhsar:93,Laughlin:98,Volovik:97,Balatsky:98}), our
model is thus of great interest.

\section{The model}
We limit our discussion to clean two spatial dimensional
superconductors in respect of the layered structure of
cuprates. Following Lee and Wen \cite{Lee-Wen:97sc}, we assume that
the elementary excitations in the superconducting state are well
defined fermionic quasiparticles  with dispersion 
$E(\Vp)=\sqrt{(\epsilon(\Vp)-\mu)^2 +\Delta^2(\Vp)}$, where $\Vp=(p_1,
p_2)$, 
$\epsilon(\Vp) =2 t_f[\cos (p_1 a) +\cos (p_2 a)]$ and 
$\Delta(\Vp) ={1\over 2}\Delta_0[\cos (p_1a) -\cos (p_2 a)]$
with $a$ the square lattice  constant. Obviously,
$\Delta(\Vp)$ is a $d_{x^2-y^2}$-wave gap. The Fermi surface consists
of  four isolated nodes at
$\Vp=(\pm{\pi\over 2a},\pm{\pi\over 2a})$ where the gap is vanishing. In
the following, we shall consider the case of chemical potential
$\mu\approx 0$ \cite{Lee-Wen:97sc,Wen-Lee:98}.

In the BCS formulation of the high $T_c$ superconductor, the above
quasiparticle spectrum can be obtained from the Bogoliubov--Nambu
Hamiltonian \cite{remark:Bogoliubov-Nambu}
\begin{equation}
{H} =\sum_{\Vp} \left(c^\dag_\uparrow(\Vp), c_\downarrow(-\Vp) \right)
 \left(\begin{array}{cc}
	 \epsilon(\Vp)  & -i\tilde{\Delta}(\Vp) \\
	i{\tilde{\Delta}}^*(\Vp) & -\epsilon(\Vp)
 \end{array}\right)
{c_\uparrow(\Vp) \choose c_\downarrow^\dag(-\Vp)}, \label{eq:HMF}
\end{equation}
where $c_\sigma(\Vp)$ and $c^\dag_\sigma(\Vp)$ are
the quasi-electron annihilation and creation operators with spin
indices $\sigma=\uparrow,\downarrow$.
$\tilde{\Delta}(\Vp)$ is related to $\Delta(\Vp)$ 
by $\tilde{\Delta}=\exp(i\phi)\Delta$ where 
$\phi$ can be interpreted as the phase of the complex superconducting
order parameter. (By definition,
$\Delta(\Vp)\equiv |\tilde{\Delta}(\Vp)|$ is real.)
Since our aim is to obtain an effective description at low energies
and long length-scales, it is sufficient to focus on the gapless modes
near the $d$-wave nodes (known as nodons
\cite{Balentz-Fisher-Nayak:98pre}), 
integrating out those electrons far away in
the Brillouin zone. Hence, it is legitimate to linearize the
quasiparticle Hamiltonian (\ref{eq:HMF}) and perform a rotation  in
momentum space. 
In the vicinity of each node, say
$({\pi\over 2a}, {\pi\over 2a})$, we have 
\begin{equation}
\begin{array}{rcl}
\epsilon(\Vp) &\rightarrow&
v_F p^\prime_2, \\
{\Delta}(\Vp) &\rightarrow&
v_\Delta p^\prime_1,
\end{array} \label{eq:Elinear}
\end{equation}
where
$p^\prime_1=(1/\sqrt{2})(-p_1+p_2)$,
$p^\prime_2=(1/\sqrt{2})(-p_1-p_2+\pi/a)$,
 $v_F=2\sqrt{2}t_fa$ and
$v_\Delta=\Delta_0 a/\sqrt{2}$.
In the coordinate space, the resulting continuum theory reads
\begin{equation}
{H} = \sum_r\int d^2 x \left(c^\dag_\uparrow(\Vx), c_\downarrow(\Vx) \right)_r
 \left(\begin{array}{cc}
	 v_F\hat{p}_2  & -i \tilde{\Delta}(\hat{\Vp})  \\
	i\tilde{\Delta}^*(\hat{\Vp}) & -v_F\hat{p}_2
 \end{array}\right)
{c_\uparrow(\Vx) \choose c_\downarrow^\dag(\Vx)}_r, \label{eq:Hx}
\end{equation}
where the index $r$ runs over four nodal points,
$\hat{p}_i\equiv -i\partial_i$ with index $i=1,2$, and
$\tilde{\Delta}(\hat{\Vp})=e^{i\phi} v_\Delta \hat{p}_1$.

The presence of an external {\it constant} magnetic $B$ field perpendicular
to the superconducting plane can be incorporated into our effective
Hamiltonian (\ref{eq:Hx}) by introducing an external (classical)
gauge field $\VA$ and replacing $\hat{\Vp} \rightarrow \hat{\Vp}
\pm e\VA$ with $\pm$ depending on whether the quasiparticle is
electron- or hole-like
\cite{Volovik:bk92,Anderson:98pre}.  
Here is a subtle point. In order to maintain the Hamiltonian 
both gauge invariant and
hermitian, we should not introduce gauge
fields through the complex superconducting order
parameter (or gap) function. That is, we replace
$\epsilon(\hat{\Vp}) \rightarrow \epsilon(\hat{\Vp}\pm e\VA)$ whereas
keeping  $\tilde{\Delta}(\hat{\Vp})$ unchanged
\cite{remark:PALee}.  Alternatively, one can understand  
this by an intuitive argument: a gap function $\tilde{\Delta}$ is by
definition expressed in terms of the paring fields and the interaction
potential and therefore should be naturally interpreted as a function
of coordinate variable $\Vx$  rather than momentum
operator $\hat{\Vp}$, while $\epsilon(\hat{\Vp})$ can be thought as a
kinematic term.  
Thus,
\begin{equation}
{H} = \sum_r\int d^2 x \left(c^\dag_\uparrow(\Vx), c_\downarrow(\Vx) \right)_r
 \left(\begin{array}{cc}
	 v_F(\hat{p}_2 + eA_2)  & -i\tilde{\Delta}(\hat{\Vp})  \\
	i\tilde{\Delta}^*(\hat{\Vp}) & -v_F(\hat{p}_2 -eA_2)
 \end{array}\right)
{c_\uparrow(\Vx) \choose c_\downarrow^\dag(\Vx)}_r . \label{eq:HxA}
\end{equation}
(In our notation, the electron has charge $-e$.)
Accordingly, let us choose a gauge:
\begin{equation}
\VA=(A_1, A_2)=(0,Bx_1) . \label{eq:gauge}
\end{equation}
The above Hamiltonian is manifestly invariant under 
a U($1$) gauge transformation $\Lambda(\Vx)$:
\begin{eqnarray}
A_i(\Vx) &\rightarrow& A_i(\Vx) +\partial_i \Lambda(\Vx), \nonumber\\
c_\sigma(\Vx) &\rightarrow& e^{-ie\Lambda(\Vx)}c_\sigma(\Vx), \\ 
\tilde{\Delta} &\rightarrow& e^{-i2e \Lambda(\Vx)} \tilde{\Delta}. \nonumber 
\end{eqnarray}
If $\tilde{\Delta}(\hat{\Vp})$ had been treated in the same way as 
$\epsilon(\hat{\Vp})$, we  would have got into trouble of
either violating the U($1$) gauge invariance or having a
non-hermitian 
Hamiltonian \cite{remark:PALee}. 

Following Anderson \cite{Anderson:98pre}, we transform away the $\phi$
dependence of  the Hamiltonian (\ref{eq:HxA}) by
a unitary transformation $H^\prime= UH U^{-1}$ with 
\begin{equation}
U=\left( \begin{array}{cc}
	e^{i\phi} & 0 \\
	0 & 1 
	\end{array}\right) . \label{eq:U}
\end{equation}
We then have 
\begin{equation}
{H^\prime} = \sum_r\int d^2 x 
\chi^\dag_r\left[ \sigma_3 v_F({\hat{p}_2 -eA_2}) +
\sigma_2v_\Delta\hat{p}_1 \right]\chi_r    \label{eq:Hchi}
\end{equation}
where $\sigma_a$'s ($a=1,2,3$) are the usual Pauli matrices and the 
fermionic fields 
$$\chi_r =U
{c_\uparrow \choose c_\downarrow^\dag}_r. 
$$
To  obtain Eq.~(\ref{eq:Hchi}), we have used the fact that
\begin{equation}
\nabla\phi = -2e{\mbf A}+ m\mbf{v}_s , \label{eq:phi-A}
\end{equation}
and assumed that, deep inside a superconductor, 
the supercurrent velocity $v_s\approx 0$
(i.e., the contribution of vortex lattice has been ignored
\cite{remark:vortex}). 
Anderson \cite{Anderson:98pre} suggested that Eq.~(\ref{eq:Hchi}) 
is the effective Hamiltonian 
appropriate for hole-like excitations in high $T_c$ superconductors. 
There exists another unitary matrix $U$ that would lead to an
effective 
Hamiltonian for electron-like excitations, identical with
(\ref{eq:Hchi}) except for $e\rightarrow -e$ (see
Ref.~\cite{Anderson:98pre} for detail).   
The Hamiltonian (\ref{eq:Hchi}) can be equivalently recast into a Dirac 
(``relativistic'') form. To achieve this, let us rewrite $H^\prime$ in
terms of  new fermion fields, related to the spinors $\chi_r$ via
a unitarity transformation,
\begin{equation}
\psi_r = {1\over \sqrt{2}}(\sigma_1+\sigma_3)\chi_r ={1\over
\sqrt{2}} {e^{i\phi}c_\uparrow + c_\downarrow^\dag \choose 
e^{i\phi}c_\uparrow - c_\downarrow^\dag }_r ,
\label{eq:psi}
\end{equation} 
and rescale our coordinate space such that $x_1 \rightarrow
(v_\Delta/v_D) x_1, x_2\rightarrow (v_F/v_D) x_2$,  where
$v_D\equiv\sqrt{v_Fv_\Delta}$ takes place  of the velocity of light in
a true relativistic quantum theory. Also, we are free to replace
$p_1\rightarrow (p_1-eA_1)$ for convenience since we are working in the
gauge (\ref{eq:gauge}) in which $A_1=0$.   Thus, the effective
Hamiltonian for nodal particles (quasiholes, for instance) reads
\begin{equation}
{H^\prime} = v_D\sum_r\int d^2 x 
{\psi}^\dag_r\gamma_3(\gamma_iD_i) \psi_r ,  \label{eq:Hpsi}
\end{equation}
where we have introduced 
the covariant derivative $D_i=\partial_i -ieA_i$ (with $i=1,2$) and the
$2\times 2$ Dirac
matrices $\gamma_\mu =(\sigma_1,\sigma_2,\sigma_3)$ which satisfy the
Clifford algebra $\{\gamma_\mu,\gamma_\nu\}= 2\delta_{\mu\nu}$
($\mu,\nu=1,2,3$). 

We are now in a position to propose a low energy effective field theory
for quasiparticles in the high $T_c$ superconducting state.  The
partition functional and Lagrangian \cite{Semenoff+:89} read
(in Euclidean space
with $\hbar\equiv v_D\equiv 1$ and $\beta=1/k_BT$) respectively
\begin{equation}
Z=\int \CD\overline{\psi}\CD\psi \ 
e^{-\int_0^\beta d\tau\int d^2 x \CL}	
\end{equation}
and 
\begin{equation}
{\CL}  = \overline{\psi}_r(\gamma_3 \partial_\tau +\gamma_iD_i)\psi_r 
-{g\over 2N_{\mathrm{f}}} (\overline{\psi}_r\psi_r)^2, 
\qquad \mbox{($i=1,2$)},
\label{eq:L[psi]}
\end{equation}
where $\overline{\psi}_r\equiv \psi^\dag_r\gamma_3$.
The $2$-component spinor $\psi_r$ represents the nodal particle (e.g.,
quasi-hole) field  with a ``flavor'' index $r=1,2,\cdots,
N_\mathrm{f}$ ($N_\mathrm{f}=4$ for the physical case of four nodes). 
(Hereafter, ``flavor'' indices $r$ shall be suppressed in our
notation.) 
In this theory, we have introduced  a four-fermion interaction
$V_{\mathrm{int}} =-{g\over 2N_{\mathrm{f}}}  (\overline{\psi}\psi)^2$
with $g$ the 
coupling constant of dimensionality 
of [mass]$^{-1}$ to mimic
the quasiparticle interactions in superconducting states. 
The theory (\ref{eq:L[psi]}) is known to particle physicists as one
version of the $U(N_{\mathrm f})$ 
Gross--Neveu model in $d=2+1$ in a constant magnetic field. 
A mass term is excluded
by parity (space-reflection invariance),
\begin{equation}
\psi(x_1,x_2,\tau) \rightarrow \gamma_1 \psi(-x_1,x_2,\tau)  \,.
\label{eq:parity}
\end{equation}  
As we shall see, it is this discrete symmetry that suffers 
breakdown. 

Similar  models
\cite{+Mavromatos:97,Semenoff+:98} were proposed to
study the magnetic catalysis phenomenon in high $T_c$ superconducting
states. However, our model is different. To see this,
one can  rewrite
the Lagrangian (\ref{eq:L[psi]}) as 
$\CL= \overline{\Psi}_l(\tilde{\gamma}_3 \partial_\tau +
\tilde{\gamma}_iD_i)\Psi_l
-{g\over 2N_{\mathrm{f}}} (\overline{\Psi}_l\tau\Psi_l)^2$ (where $l=1,2$) by
combining two $2$-component fermions in a new four-component spinor
$\Psi_l$ and introducing $4\times 4$ 
matrices $\tilde{\gamma}_\mu=\mathrm{diag}(\sigma_\mu,-\sigma_\mu) $
and $\tau=\mathrm{diag}(1,-1)$. Comparing our theory with 
those of Refs.~\cite{+Mavromatos:97} and 
\cite{Semenoff+:98} shows that we have chosen  a different
four-fermion interaction. Our choice, which shall lead to a
mass term that breaks parity, is motivated by the popular studies of
parity-breaking in high $T_c$ superconductors
\cite{Rokhsar:93,Laughlin:98,Volovik:97,Balatsky:98}.

\section{Free energy and gap equation}

The quartic term in Eq.~(\ref{eq:L[psi]}) can be canceled by
adding an expression that is quadratic in an auxiliary field
$\sigma(\Vx,\tau)$, and that vanishes when $\sigma$ is integrated out. This
results in the replacement of Eq.~(\ref{eq:L[psi]}) with the
equivalent Lagrangian
\begin{equation}
\CL=\overline{\psi}(\gamma_3\partial_\tau +\gamma_iD_i)\psi 
+\sigma (\overline{\psi}\psi) + {N_{\mathrm{f}}\over 2g}\sigma^2 .
\label{eq:L[psi,sigma]}
\end{equation}
Apparently, $\CL$ has a stationary point at
$\sigma= -{g\over N_{\mathrm{f}}} 
(\overline{\psi}\psi)$.
Now, the partition functional is written as
\begin{equation}
Z=\int \CD\overline{\psi}\CD\psi\CD\sigma \ 
e^{-\int_0^\beta d\tau\int d^2 x \CL}	.	\label{eq:Z2}
\end{equation}

To leading order in $1/N_{\mathrm{f}}$ \cite{remark:1/N}, the
effective action $\Gamma[\sigma]$ is defined by
\begin{equation}
e^{-\Gamma[\sigma]}=\int \CD\overline{\psi}\CD\psi  
e^{-\int_0^\beta d\tau\int d^2 x \CL}.
\end{equation}
The Lagrangian (\ref{eq:L[psi,sigma]}) 
is quadratic in fermion fields and we can formally
integrate out them. Thus, 
\begin{equation}
\Gamma[\sigma]=\beta \CV_2{N_{\mathrm{f}}\over 2g}\sigma^2 
-\ln(\det K) ,
\label{eq:Gamma} 
\end{equation}
where $\CV_2=\int d^2 x$ is the area of the space and the ``matrix''
\begin{eqnarray}
K_{r\Vx\tau,r^\prime\Vx^\prime\tau^\prime} &=& 
-[\gamma_iD_i  +\gamma_3\partial_\tau +\sigma]
\delta^2(\Vx-\Vx^\prime) \delta(\tau-\tau^\prime)\delta_{rr^\prime} .  
\label{eq:Kmatrix}
\end{eqnarray}
We only have even powers of $\sigma$ because the trace of an odd
number of Dirac matrices vanishes (alternatively, because $\sigma$
changes sign under parity transformation
(\ref{eq:parity})). At first glance, it would
seem that $\avg{\sigma}=0$ is automatically a stationary point since
$\Gamma$ is even in $\sigma$. However, if we are interested in the
breakdown of parity, the issue is precisely whether there
are stationary points other than $\avg{\sigma}=0$. Fortunately, to
settle this issue, we only need to compute $\Gamma$ for constant
$\sigma$ \cite{Coleman:bk85:ch8}. 

Since $K$ is invariant under imaginary time
translations, it is convenient to perform a Fourier transformation on
$\tau$:
\begin{eqnarray}
K_{r\Vx n,r^\prime\Vx^\prime n^\prime}&=&\int_0^\beta {d\tau\over \sqrt{\beta}}
e^{i\omega_n \tau} \int_0^\beta {d\tau^\prime\over \sqrt{\beta}}
e^{-i\omega_{n^\prime} \tau^\prime}
K_{r\Vx\tau,r^\prime\Vx^\prime\tau^\prime} \nonumber \\
&=&-[\gamma_iD_i -i\omega_n\gamma_3 +\sigma] \delta^2(\Vx-\Vx^\prime)
\delta_{n n^\prime} \delta_{rr^\prime},
\end{eqnarray} 
where $\omega_n=(2n+1)\pi/\beta$ is the usual Matsubara frequency for
fermions.
Then, we have
\begin{eqnarray}
\ln(\det K) 
&=& N_{\mathrm{f}}\int d^2 x \sum_{n=-\infty}^{+\infty}
 \tr\left\{ \avg{\Vx| \ln(\gamma_iD_i 
-i\omega_n\gamma_3 +\sigma) |\Vx} \right\} \nonumber \\
&=&{1\over 2} N_{\mathrm{f}}\int d^2 x 
\sum_{n=-\infty}^{+\infty} \tr\left\{ \avg{\Vx| \ln\left(
(i\gamma_iD_i)^2+\omega_n^2 +\sigma^2\right) |\Vx}
\right\} \label{eq:Kmatrix2}
\end{eqnarray}
in virtue of  the fact that
\begin{eqnarray}
&&
\sum_{n} \tr\ln\left(\gamma_iD_i -i\omega_n\gamma_3 +
\sigma\right) 
= \sum_{n} \tr\ln \left( 
\gamma_3[\gamma_iD_i -i\omega_n\gamma_3 + \sigma]\gamma_3\right)
\nonumber \\
&=&\sum_{n} \tr\ln\left(-\gamma_iD_i
+i\omega_{-(n+1)} \gamma_3 + \sigma\right)
=\sum_{n} \tr\ln\left(-[\gamma_iD_i -i\omega_n\gamma_3] +
\sigma\right) .
\end{eqnarray}
Therefore, $\ln(\det K)$ can be expressed in terms of an 
integral over the proper time $s$,
\begin{equation}
\ln(\det K) = -{N_{\mathrm{f}}\over 2}\sum_n\int d^2 x \int_0^\infty
{ds\over s} \tr \avg{\Vx|e^{-is[(i\gamma_iD_i)^2+
\omega_n^2+\sigma^2]}|\Vx}, 	\label{eq:Kmatrix3}
\end{equation}
where $(i\gamma_iD_i)^2=-D_iD_i+ieB\gamma_1\gamma_2$.
Following Schwinger's proper time approach \cite{Schwinger:51proper}, 
we find the matrix element
\begin{equation}
\avg{\Vx^\prime|e^{-is(i\gamma_iD_i)^2}|\Vx^{\prime\prime}} = 
{-i\over 4\pi}R(\Vx^\prime,\Vx^{\prime\prime};s) eB[\cot(eBs)
+i\gamma_3],  \label{eq:me}
\end{equation}
where 
$$R(\Vx^\prime,\Vx^{\prime\prime};s)=\exp[
{i\over 4}(\Vx^\prime-\Vx^{\prime\prime})^2 eB\cot(eBs) 
+ie \int_{\Vx^{\prime\prime}}^{\Vx^\prime} dx_i A_i (\Vx)],
$$
in which the integration path is  a straight line.
Substituting Eq.~(\ref{eq:me}) into Eq.~(\ref{eq:Kmatrix3}), we get
\begin{equation}
\ln(\det K)= {iN_{\mathrm{f}}eB\over 4\pi} 
\sum_n\int d^2 x \int_0^\infty
{ds\over s} e^{-is(\omega_n^2+\sigma^2)} \cot(eBs) .	
\label{eq:Kmatrix4}
\end{equation}
Inserting Eq.~(\ref{eq:Kmatrix4}) in Eq.~(\ref{eq:Gamma}) gives us the
effective action. Let $\Gamma[\sigma]\equiv \beta\CV_2 F(\sigma)$
where $F(\sigma)$ is called the free energy. After 
a Wick rotation $s\rightarrow -is$,
\begin{equation}
F(\sigma)={N_{\mathrm{f}}\sigma^2\over 2g} +{N_{\mathrm{f}} eB\over 4\pi\beta}
\sum_n\int_{1/\Lambda^2}^\infty {ds\over s}
e^{-s(\omega_n^2+\sigma^2)} \coth(eBs) , 	\label{eq:FreeE}    
\end{equation}
where we have explicitly introduced the ultraviolet cutoff (Debye
frequency) $\Lambda$.
An integral very analogous to that of Eq.~(\ref{eq:FreeE}) was
evaluated in Appendix B of Ref.~\cite{Gusynin+:94+95}. We thus skip all
intermediate 
analyses and directly go the final result of the free energy,
\begin{eqnarray}
F(\sigma) &=& {N_{\mathrm{f}}\over 2\pi}\left[{1\over 2}M_s\sigma^2
-{\sqrt{2}\over l^3}\zeta(-{1\over 2},{\sigma^2 l^2\over 2}+1)
-{\sigma\over2l^2} \right]
\nonumber\\
&& -{N_{\mathrm{f}}\over2\pi\beta l^2}\left[ \ln(1+e^{-\beta |\sigma|}) +
2\sum_{k=1}^\infty \ln(1+e^{-\beta\sqrt{\sigma^2+ 2k/l^2}}) \right]
+O({1\over \Lambda}),
\label{eq:FreeEfinal}
\end{eqnarray}
where $l=|eB|^{-1/2}$ is the magnetic length, $M_0={\pi\over g}
-{\Lambda\over\sqrt{\pi}}$ is of the dimension of mass, and $\zeta(z,q)$ is
the generalized Riemann Zeta function
\cite{Gradshteyn-Ryzhik:80}.  This result is exact for arbitrary
strength of $B$.

The gap equation, $\partial F/\partial\sigma=0$, reads
\begin{eqnarray}
0 &=& \sigma\left[ M_0-{1\over\sqrt{2}l} 
\zeta({1\over 2},{\sigma^2 l^2\over 2}+1) -{1\over
2l^2\sigma}\tanh{\beta|\sigma|\over 2} 
+{2\over
l^2}\sum_{k=1}^{\infty}{(\sigma^2+{2k/ l^2})^{-1/2} \over
(e^{\beta\sqrt{\sigma^2+2k/l^2}}+1)}
\right].	\label{eq:gap}
\end{eqnarray}
The nontrivial solution to Eq.~(\ref{eq:gap}) yields the quantum
thermal average $\avg{\sigma}$.  As
Eq.~(\ref{eq:L[psi,sigma]}) implies, a non-vanishing $\avg{\sigma}$
means that fermions gain dynamical mass, $m=\avg{\sigma}$. In other
words, a gap opens at the  nodes of the quasiparticle spectrum.  
Meanwhile, parity
is  broken by the mass term $\avg{\sigma}
(\overline{\psi}\psi)$ to appear  in
the effective Lagrangian.
As $B\rightarrow 0$, Eq.~(\ref{eq:gap}) can be explicitly solved,
yielding the known result \cite{Rosenstein-Warr-Park:91}
\begin{equation}\textstyle
\avg{\sigma}=-M_0+2\beta^{-1} \ln({1\over 2}[1+\sqrt{1-4\exp(\beta
M_0)}]). \label{eq:gapB=0}
\end{equation}
Therefore, for some range of parameter space $(T, M_0)$  there is
spontaneous parity breakdown.
Obviously, as $T\rightarrow 0$, Eq.~(\ref{eq:gapB=0}) has a nontrivial
solution only if $M_0<0$ (i.e., $g>\pi^{3/2}/\Lambda$). 
For finite $B$,
the magnetic field however changes the situation
dramatically. Firstly, an external magnetic field itself explicitly breaks
the parity invariance of the Lagrangian (\ref{eq:L[psi]}) since it
requires that the Lagrangian be invariant under transformation
(\ref{eq:parity}) together with transformations $(A_1, A_2) \rightarrow
(-A_1, A_2)$ and $B\rightarrow -B$.  Secondly, 
as we learned from Gusynin {\it et al.}
\cite{Gusynin+:94+95}, the magnetic field can catalyze parity
breaking. This shall be discussed further in the following section.

\section{Comparison with experiment}

Next, we shall investigate the free energy (\ref{eq:FreeEfinal}) and the gap
equation (\ref{eq:gap}) numerically for the experiment
of Krishana {\it et al.}\cite{Krishana+:97}.
The theory has two free parameters, 
$g$ and $v_D$. The latter determines the typical
energy scale of the theory.
 (In this paragraph only, we shall explicitly restore
$\hbar$ and $v_D$ in order to discuss experiments.) 
The lattice constant for the cuprate BSCCO 
is $a\sim 5.41$\AA (see, for
example, \cite{Poole-Farach-Creswick:bk95:p198}).  
The point-contact tunneling experiment on BSCCO
\cite{Ekino+:89} shows that the ratio of $2\Delta_0/k_BT_c$ is roughly
$11.6\sim 12.4$. ($T_c$ denotes the superconducting
critical temperature, and $\Delta_0$ is the gap associated with the
occurrence of superconductivity and needs not to be confused with the
gap given by Eq.~(\ref{eq:gap}).) 
If we take $12.4$ as  the ratio and 
$T_c=92K$ \cite{Krishana+:97}, then 
$\Delta_0\approx 49.15$meV. This gives
$\hbar v_2=\Delta_0 a/\sqrt{2} \approx 0.188$eV$\cdot$\AA. On the other hand,
the superconducting coherence length for BSCCO can be taken typically
as $\xi_0=20$\AA \cite{Balatsky:98}. By definition, $\hbar v_F
=\pi\Delta_0\xi_0 \approx
3.09$eV$\cdot$\AA. Therefore, we find
$\hbar v_D=\hbar \sqrt{v_Fv_2}
\approx 0.762$eV$\cdot$\AA. The Debye cutoff $\Lambda$ is of the
order of $\hbar v_D /a \sim 0.14$eV, and should be regarded as the
highest energy scale in our effective field theory. 

\begin{figure}[htbp]
\begin{center}
\epsfig{file=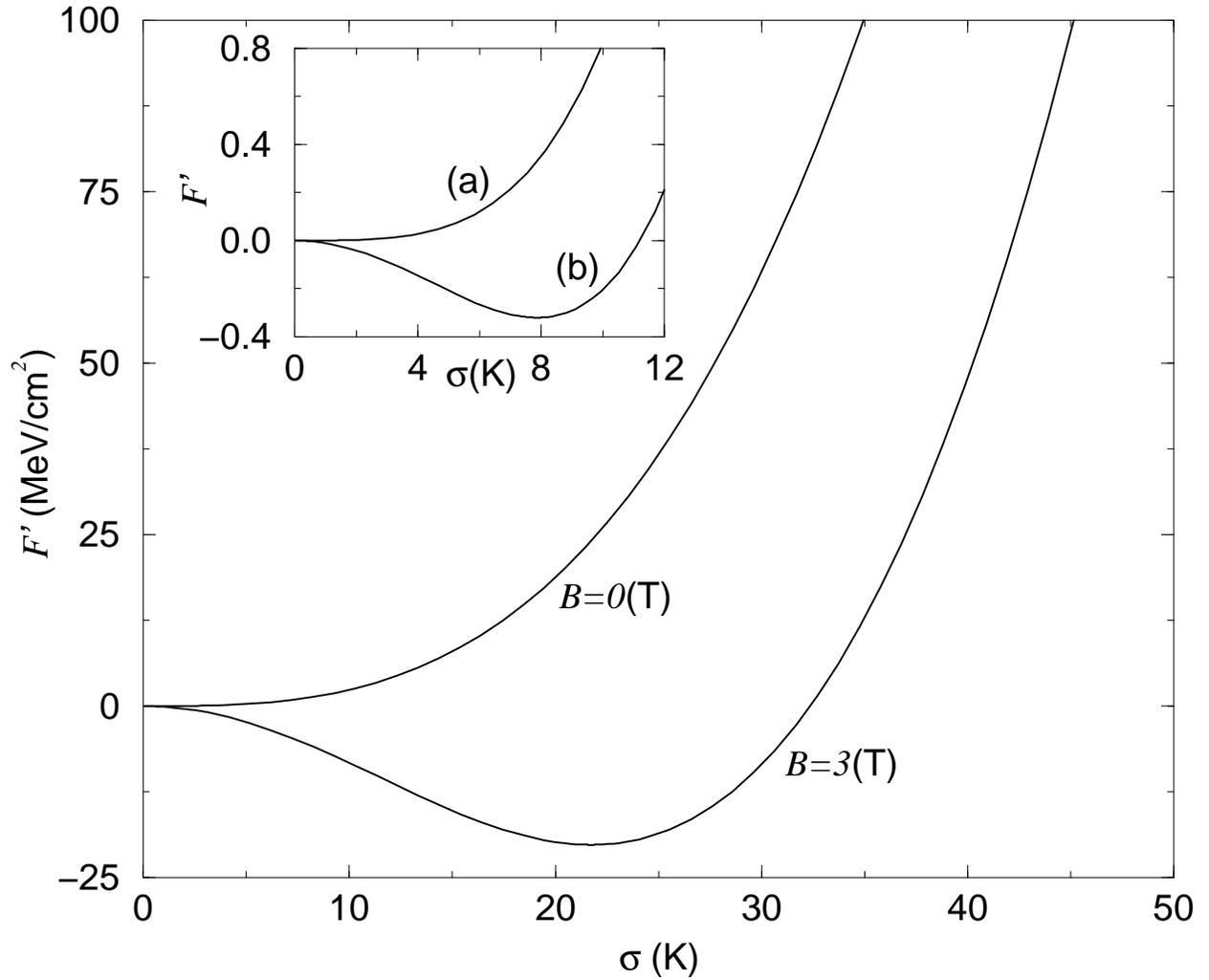,width=\linewidth}
\end{center}
\caption{The free energy $F^\prime \equiv F(\sigma) -F(0)$
as a function of $\sigma$ at $T=10$K, $M_0=0$, and 
$\hbar v_D=0.762$eV$\cdot$\AA. 
Insert: (a) $B=1.44$T is just 
below the critical point; and 
(b) $B=1.64$T slightly above the critical point. }
\label{fig:freeE}
\end{figure}

\begin{figure}[htbp]
\begin{center}
\epsfig{file=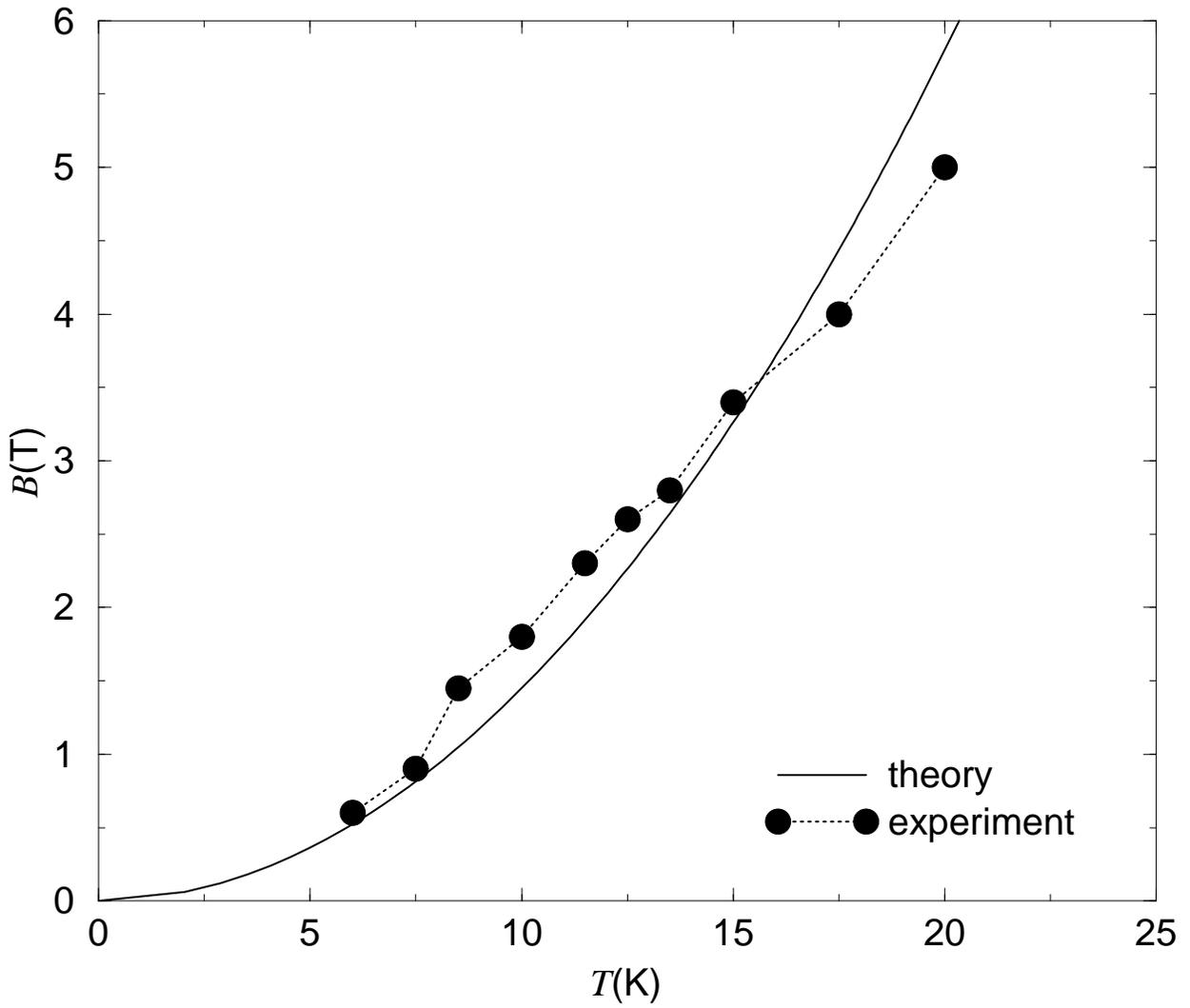,width=\linewidth}
\end{center}
\caption{The critical line of $B$--$T$ with parameters $M_0=0$ and
$\hbar v_D=0.762$eV$\cdot$\AA. The experimental data is the courtesy
of Krishana {\it et al.} \protect\cite{Krishana+:97}}
\label{fig:B-T}
\end{figure}

The free energy (\ref{eq:FreeEfinal}) is plotted in
Fig.~\ref{fig:freeE}. It exhibits that at low temperatures  a
magnetic field induces a (continuous) phase transition: $F(\sigma)$
can have a stationary point  $\avg{\sigma}\neq 0$, which
infers 
that  parity is spontaneously broken. As shown in the insert of
Fig.~\ref{fig:freeE}, a slight increase of the field  above
the critical point ($\delta B=0.2$T)
results in a gap $2\avg{\sigma}\simeq 16$K at $T=10$K. It follows that 
the quasiparticle density  is substantially suppressed by a factor
$\exp(-2\beta \avg{\sigma}) \approx 20\%$.    
Fig.~\ref{fig:B-T} shows the critical line of $B$--$T$
determined by Eq.~(\ref{eq:gap}) in comparison with the experimental
data from Krishana {\it et al.} \cite{Krishana+:97}. We find that
the experimental data  can be well fitted 
by a very small  $M_0$ \cite{remark:M_0}.  According to our theory, the
critical line means  that
parity is restored above some temperature that
depends on $B$. We thus conclude that the basic feature of the
experiment \cite{Krishana+:97} can be well explained by the present
theory. 

\section*{Acknowledgement}

It is a great pleasure to thank
my advisor Professor Steven Weinberg for encouraging
me and  reading this manuscript. Also, I would like to thank 
Jacques Distler, Willy Fischler, Patrick A. Lee, Igor Shovkovy and Grisha
Volovik for very enlightening 
discussion and criticism,  and 
K. Krishana for helpful instructions on experiments. 
This work is supported in part by NSF grant PHY-9511632 and the
Robert A. Welch Foundation.

\bibliographystyle{prsty} 
\bibliography{parity,book,paper}

\begin{thebibliography}{10}
\bibitem{remark:vortex}
For high temperature superconductor, this means that we are interesting in the
  regime $B\ll H_{c2}$ where vortex cores comprise a very small relative volume
  $\propto {B\over H_{c2}}$.
\bibitem{remark:1/N}
According to the usual prescription, we have to replace $\sigma$ in
  Eq.~(\ref{eq:Z2}) with the sum $\sigma(x) +\sigma^\prime(x)$ and integrate
  out the quantum perturbation $\sigma^\prime(x)$, including only
  one-particle-irreducible graphs. A standard power-counting argument shows
  that the contribution of the quantum perturbation to $\Gamma[\sigma]$ is
  suppressed by the power of $1/N_{\mathrm f}$.
\bibitem{remark:LandauLevel}
The Hamiltonian (\ref{eq:Hpsi}) is obviously of a ``relativistic" form, which
  is known to have a spectrum of eigenvalues $E_n= \hbar v_D\sqrt{eBn/\hbar}$
  with $n=0,1,2,\cdots$. By contrast, a non-relativistic $2$D free electron gas
  in the presence of a constant magnetic field has Landau levels $E_n=\hbar
  ({eB\over m})(n+{1\over2})$.
\end{thebibliography}

\end{document}